\documentclass[fleqn,usenatbib]{mnras}

\usepackage{newtxtext,newtxmath}

\usepackage[T1]{fontenc}

\DeclareRobustCommand{\VAN}[3]{#2}
\let\VANthebibliography\thebibliography
\def\thebibliography{\DeclareRobustCommand{\VAN}[3]{##3}\VANthebibliography}

\usepackage{graphicx}	
\usepackage{amsmath}	

\title[Compaion of V1787 Ori]{Discovery of an M-type Companion to the Herbig Ae Star V1787 Ori}

\author[R. Arun et al.]{
R. Arun,$^{1}$\thanks{E-mail: arun.roy@res.christuniversity.in}
Blesson Mathew,$^{1}$
Sridharan Rengaswamy,$^{2}$
P. Manoj,$^{3}$
Mayank Narang,$^{3}$
\newauthor
Sreeja S. Kartha,$^{1}$ and
G. Maheswar ,$^{2}$
\\
$^{1}$Department of Physics and Electronics, CHRIST (Deemed to be University), Bangalore 560029, India\\
$^{2}$Indian Institute of Astrophysics, Sarjapur Road, Koramangala, Bangalore 560034, India\\
$^{3}$Tata Institute of Fundamental Research, Homi Bhabha Road, Mumbai 400005, India
}

\date{Accepted XXX. Received YYY; in original form ZZZ}

\pubyear{2020}

\begin{document}
\label{firstpage}
\pagerange{\pageref{firstpage}--\pageref{lastpage}}
\maketitle

\begin{abstract}
The intermediate mass Herbig Ae star V1787 Ori is a member of the L1641 star-forming region in the Orion A molecular cloud. We report the detection of an M-type companion to V1787 Ori at a projected separation of 6.66$\arcsec$ (corresponding to 2577 au), from the analysis of VLT/NACO adaptive optics K\textsubscript{s}-band image. Using astrometric data from Gaia DR2, we show that V1787 Ori A and B share similar distance ($d$ $\sim$ 387 pc) and proper motion, indicating that they are physically associated. We estimate the spectral type of V1787 Ori B to be M5 $\pm$ 2 from color--spectral type calibration tables and template matching using SpeX spectral library. By fitting PARSEC models in the Pan-STARRS color-magnitude diagram, we find that V1787 Ori B has an age of 8.1$^{+1.7}_{-1.5}$ Myr and a mass of 0.39$^{+0.02}_{-0.05}$ M\textsubscript{\(\odot\)}. We show that V1787 Ori is a pre-main sequence wide binary system with a mass ratio of 0.23. Such a low mass ratio system is rarely identified in Herbig Ae/Be binary systems. We conclude this work with a discussion on possible mechanisms for the formation of V1787 Ori wide binary system.
\end{abstract}




\section{Introduction}

Multiplicity is a phenomenon generally seen in stars of all masses and evolutionary phases. The multiplicity of young stars in star formation regions are higher than that of field stars \citep{Reipurth1993, Duchene2013, Tobin2016ApJ...818...73T}. Thus, the multiplicity is directly linked to the initial star formation process \citep{Tohline2002ARA&A..40..349T}. The separation between the components of a binary system is still a topic of discussion as close binaries are of the scales of 1 - 10 au and wide binaries extent up to parsec scales (\citealp{Esteban2019} and references within). The increased technological advancements in the field of observational astronomy such as adaptive optics (AO) \citep{Davies_A0_2012, Rigaut_AO_2015PASP..127.1197R} have helped in discovering fainter binary candidates\citep{Giulia2020}. The high precision data from Gaia mission \citep{gaia2016,gAIANEW2018} has picked up more distant companions using its astrometric capabilities \citep{Esteban2019, Pittordis2019MNRAS.488.4740P}. 

In the formulation of star formation theories, it is important to address the role of multiplicity in stars, since most of the stars, in the field \citep{Duquennoy1991A&A...248..485D} or part of star-forming regions \citep{Ghez1997} are identified as binaries or higher-order systems. Earlier studies have proposed that almost 100\% of stars are born with one or more companions \citep{Abt1983}. Recent studies that use advanced astronomical telescopes and instruments estimated the multiplicity fraction (MF) of main sequence stars in various mass ranges. The studies on G dwarf stars by \cite{Raghavan2010} showed their MF to be 44 $\pm$ 2\%. The lower mass K$-$M dwarfs are reported to have MF of 26 $\pm$ 3\% \citep{Leinert1997,Delfosse2004ASPC..318..166D}. And the high mass stars with higher MF $>$80\% \citep{Sana2011IAUS..272..474S,Chini2012MNRAS.424.1925C,Duchene2013}.

Even though the studies on the main sequence binaries provide valuable information on the subject, probing earlier conditions of binarity such as in the case of the pre-main sequence (PMS) phase could provide a broader idea on the aspect of binary system formation \citep{Mathieu1994}. The low mass PMS stars are known as T Tauri stars \citep{Joy1945ApJ...102..168J} and their intermediate mass counterparts are Herbig Ae/Be (HAeBe) stars \citep{Herbig1960ApJS....4..337H}. PMS binary systems can be better understood by studying the environments of T Tauri and HAeBe stars \citep{Mathieu1994}. The studies on the multiplicity of T Tauri stars are numerous and started decades ago \citep{Ghez1993, Brandner1996}. The companion frequency of stars of mass 0.25 $-$ 2.5 M\textsubscript{\(\odot\)} in the Taurus region is in between 2/3 $-$ 3/4, suggesting a MF in the range 65$-$75\% \citep{Kraus2011}. However, multiplicity studies of HAeBe stars are less explored, although the initial studies of \cite{Reipurth1993} and \cite{Leinert1997} are worth mentioning. Specific analysis on the binary fraction of HAeBe stars was conducted by \cite{Baines2006} and \cite{Wheel2010}, who found MF to be 68\% and 74\%, respectively. It may be noted that the binary studies in HAeBe stars are affected by small number statistics and hence the inclusion of new members will provide a better understanding of the binary population in the PMS phase. Also, clarity about the mass ratio and the formation mechanisms of PMS binaries can be brought out from the detection of new PMS binaries such as V1787 Ori. 


\begin{figure*}
   \centering
   \includegraphics[width=0.9\textwidth,height=0.7\textwidth]{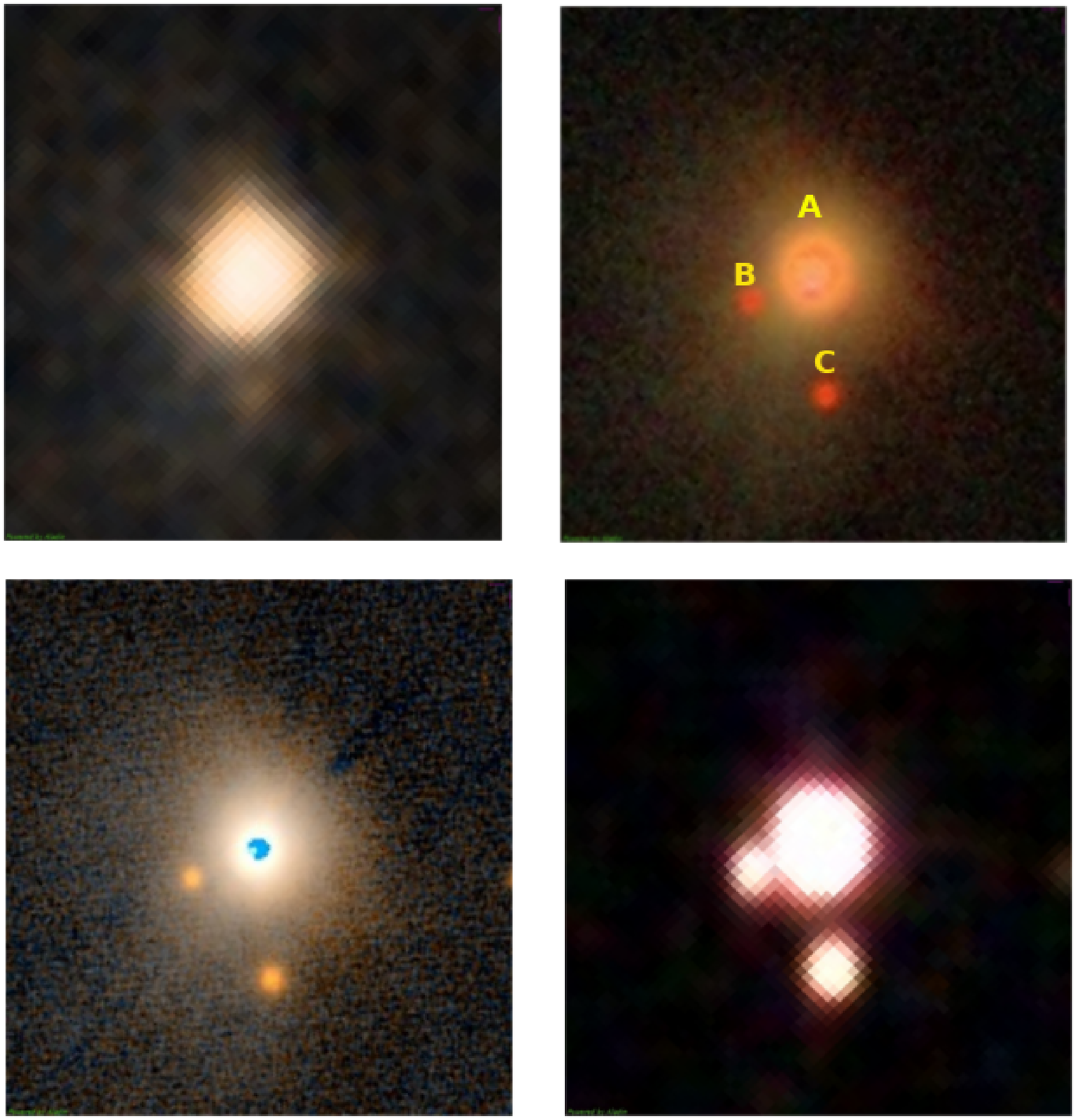}
   \caption{Figure shows RGB images of the region around V1787 Ori (58.52"x 34.13") from DSS (upper left), SDSS (upper right), Pan-STARRS (lower left), and 2MASS (lower right) surveys. The DSS image shows only the HAeBe star V1787 Ori. The objects marked as "A", "B" and "C" in the SDSS image are V1787 Ori, 2MASS J05380972-0649189 and 2MASS J05380922-0649272, respectively. }
   \label{Fig1}
   \end{figure*}
   
\subsubsection*{About the object}
V1787 Ori (Parenago 2649) is a HAeBe star of spectral type A5 \citep{The1994, VIEIRA2003}. This star belongs to L1641 cloud of the Orion A molecular cloud complex \citep{Nakajima1986}. V1787 Ori was first identified in the survey by \cite{Parenago1954} and its PMS nature was established by \cite{Weintraub1990}. \citet{Arun2019} and \citet{Vioque2018} estimated the stellar parameters such as distance, mass, and age of V1787 Ori from the analysis of data from Gaia Data Release 2 (Gaia DR2). The mass and age estimated by \cite{Arun2019} are 2.04$^{+0.03}_{-0.02}$ M\textsubscript{\(\odot\)} and 0.57$^{+0.02}_{-0.02}$ Myr, respectively, whereas \cite{Vioque2018} found the values to be 1.659$^{+0.094}_{-0.083}$ M\textsubscript{\(\odot\)} and 7.4$^{+0.6}_{-1}$ Myr, respectively. 
The distinct difference in age is due to the difference in total-to-selective extinction (R$_V$) values used in these studies (More details in Sect. \ref{sec:age}).  

During their study, \citet{Arun2019} noticed two stars at the proximity of V1787 Ori, within an $11\arcsec$ search radius. In this study, with the available photometric and astrometric data, we investigate whether they are associated with V1787 Ori. The rest of this paper is organized as follows. The data used for this study is described in Sect. \ref{sect:Obs}. Sect. \ref{sec:res} explains the different analyses carried out to estimate the stellar parameters of the V1787 Ori binary system. A discussion about the identified companion of V1787 Ori and the possible binary formation scenarios are provided in Sect. \ref{sec:disc}, and major results are summarized in Sect. \ref{sect:conclusion}.  

\section{Data inventory}
\label{sect:Obs}

 The multi-wavelength photometric data of the three stars are available from various surveys such as SDSS DR12 \citep{SDSSPHOT2015A}, 2MASS \citep{Skrutskie2006}, Pan-STARRS DR1 \citep{Chambers2016}, Gaia DR2 \citep{GAIAPHOT2018}, and VISION-VISTA \citep{Meingast2016}. The parallax and proper motion of the candidate stars are taken from Gaia DR2 release \citep{gAIANEW2018}, and the distance estimates are extracted from the catalog of \cite{bailerjohns2018}. The VLT NAOS-CONICA (NACO) K\textsubscript{s} imaging data (PIs: M. Petr-Gotzens and G. Dûchene \& Program ID-080.C-0811(A)) of the object field is available in the European Southern Observatory (ESO) archive. The observations taken at two different epochs in the object field are unpublished and were recorded on the nights of 25-02-2008 and 22-03-2008 in the imaging mode, with an integration time of 1.5~s. We used these datasets along with the theoretical isochrones, evolutionary tracks, and spectral templates to study the properties of the V1787 Ori system.

\begin{table*}
\centering
\caption{Table provides the Gaia DR2 astrometric parameters of V1787 Ori A and the two nearby stars. The distance estimates are taken from \protect\cite{bailerjohns2018}.}
\label{tab:my-table1}
\begin{tabular}{|c|c|c|c|c|}
\hline
Object & \begin{tabular}[c]{@{}c@{}}d\\ (pc)\end{tabular} & \begin{tabular}[c]{@{}c@{}}Parallax \\ (mas)\end{tabular} & \begin{tabular}[c]{@{}c@{}}$\mu_{\alpha *}$\\ (mas yr\textsuperscript{-1})\end{tabular} & \begin{tabular}[c]{@{}c@{}}$\mu_{\delta}$\\ (mas yr\textsuperscript{-1})\end{tabular} \\ \hline
V1787 Ori A & 387$^{+8}_{-8}$ & 2.554$\pm$0.0531 & 0.243$\pm$0.09 & -0.538$\pm$0.09 \\ 
V1787 Ori B (2MASS J05380972-0649189) & 371$^{+67}_{-49}$ & 2.7419$\pm$0.3826 & -0.254$\pm$0.61 & -0.197$\pm$0.67 \\ 
2MASS J05380922-0649272 & 423$^{+66}_{-51}$ & 2.3815$\pm$0.3027 & -2.165$\pm$0.55 & 12.72$\pm$0.59 \\ \hline
\end{tabular}%
\end{table*}

\section{Results}\label{sec:res}
The sources of interest around V1787 Ori are illustrated in \autoref{Fig1}. The sources are found adjacent to V1787 Ori in SDSS, PanSTARRS DR1, and 2MASS images, whereas the system is not resolved well in the DSS color composite. The star at a separation of $6.6\arcsec$ from V1787 Ori is 2MASS J05380972-0649189, and that at $11\arcsec$ separation is identified as 2MASS J05380922-0649272. In the following sections, we analyze the stellar properties of these sources in order to assess their companionship with V1787 Ori.

\subsection{Gaia astrometry analysis}

The Gaia mission is launched by the European Space Agency (ESA) on 19th December 2013 and is still in operation. Gaia, with its two data releases, gives high precision astrometric and photometric measurements of about 1.3 billion stars in our Galaxy. For this work, we have used the parallax and proper motion values from Gaia DR2 release \citep{gaia2016,gAIANEW2018}. 

\subsubsection{Distance measurements from Gaia DR2}

To see whether the two nearby stars, 2MASS J05380972-0649189 and 2MASS J05380922-0649272 are associated with V1787 Ori, we analyze the distance and proper motion values of all the three stars. Gaia DR2 data for the central star V1787 Ori and two nearby stars are extracted using the J2000 epoch coordinates of V1787 Ori with a $12\arcsec$ search radius in the Vizier catalog services. The mean distance ($d$) values and the corresponding upper and lower bounds of the three stars are taken from the catalog of \cite{bailerjohns2018}. The $d$ values for V1787 Ori, 2MASS J05380972-0649189 and 2MASS J05380922-0649272 are 387$^{+8}_{-8}$ pc, 371$^{+67}_{-49}$ pc and 423$^{+66}_{-51}$ pc, respectively. 2MASS J05380972-0649189 is at a similar distance, within uncertainties, with that of V1787 Ori. However, in order to see whether they are bound, we need to analyze the proper motion values as well. 

\subsubsection{Analysis of proper motion values}

We made use of Gaia DR2 astrometric data such as proper motion values in right ascension ($\mu_{\alpha *}$ = $\mu_{\alpha}cos\delta$) and declination ($\mu_{\delta}$) to confirm the association of the two nearby stars with V1787 Ori. The astrometric quality of Gaia DR2 is quantified by the re-normalized unit weight error (RUWE) parameter. The parameter gives the goodness-of-fit statistic of the astrometric solution of each source observed in the Gaia DR2. For good quality astrometry for a star, the criterion of RUWE < 1.4 should be satisfied \citep{Lindegren2018}. The three stars that are under consideration in this study satisfy this RUWE criterion. \autoref{tab:my-table1} lists the distance and proper motion values of all the three stars. The $\mu_{\alpha *}$ and $\mu_{\delta}$ values of of V1787 Ori are 0.243$\pm$0.09 mas yr\textsuperscript{-1} and -0.538$\pm$0.09 mas yr\textsuperscript{-1}, respectively. These values are within uncertainties similar to the proper motion values for 2MASS J05380972-0649189, which is given as -0.254$\pm$0.61 mas yr\textsuperscript{-1} and -0.197$\pm$0.67 mas yr\textsuperscript{-1}, respectively. The proper motion values of the third star, 2MASS J05380922-0649272 are -2.165$\pm$0.55 mas yr\textsuperscript{-1} and 12.72$\pm$0.59 mas yr\textsuperscript{-1}, respectively, which is completely distinct from V1787 Ori and 2MASS J05380972-0649189. Hence, due to the similarity in the distance and proper motion values, we suggest that 2MASS J05380972-0649189 (hereafter V1787 Ori B) could be a binary companion to V1787 Ori (hereafter V1787 Ori A).

Although the parallax and proper motions of V1787 Ori A \& B are similar, the uncertainties in proper motion are considerably high for V1787 Ori B. Thus, we need to assess how this high uncertainty can affect our proposition that V1787 Ori is a binary system. The large uncertainty in the astrometric measurements can be due to the presence of a very close companion. However, we made sure that the astrometric quality parameter, RUWE, is less than 1.4 for V1787 Ori B, which indicates good quality astrometry and it is not contaminated by the nearby star. Also, another factor affecting the uncertainties in astrometry is the brightness of the star itself. The G magnitude of V1787 Ori B is 18.92 mag, which is $\sim$ 6 magnitudes fainter than V1787 Ori A. The typical uncertainty in the proper motion for stars with G mag in range 17 $-$ 20 mag is 0.2 $-$ 1.2 mas yr\textsuperscript{-1} (Gaia Data Release 2 - Documentation release 1.2)\footnote{https://gea.esac.esa.int/archive/documentation/GDR2/}. The proper motion uncertainties are around 0.6 mas yr\textsuperscript{-1} in both $\mu_{\alpha *}$ and $\mu_{\delta}$ for V1787 Ori B. This value is well within the optimal range mentioned in the Gaia DR2 release documentation and hence can be used for this study. We searched in literature and found that the detection of a third companion to the GQ Lup system has similar issues of large proper motion uncertainties \citep{GQLup2020}. The GQ Lup system has similarities with the V1787 Ori system in terms of projected separation between the members ($\sim$ 2500 pc) and high proper motion uncertainties. Hence, from this analysis, we suggest that the high uncertainties in proper motion values cannot disprove the association of 2MASS J05380972-0649189 to V1787 Ori.

\subsubsection{Membership of V1787 Ori System in L1641}\label{sec:vpd}

\begin{figure}
   \centering
   \includegraphics[width=1\columnwidth]{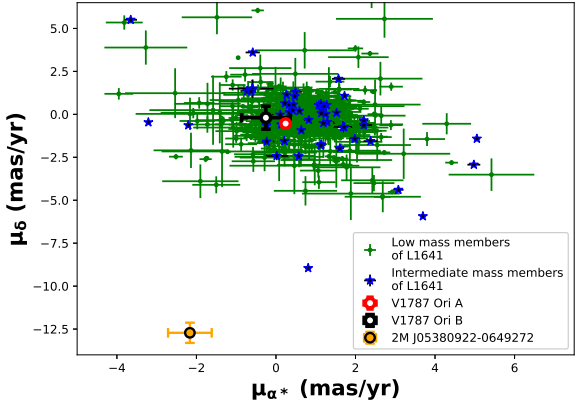}
   \caption{The figure shows the vector point diagram of members in L1641 cloud. The green dots and blue stars symbols are low mass and intermediate mass members of L1641 cloud, respectively. V1787 Ori A, B and 2MASS J05380922-0649272 are also represented.}
   \label{Fig2}
   \end{figure}

Another method we employed to test the binarity of the V1787 Ori system is from the analysis of proper motion values of all the members in the Lynd Cloud L1641. It has been shown by \citet{Nakajima1986} and \citet{Hsu2013ApJ...764..114H} that V1787 Ori A is associated with L1641 cloud. If indeed V1787 Ori B is bound to V1787 Ori A, it should also be a member of L1641. However, V1787 Ori B is not reported to be a member of the L1641 cloud. We extracted the proper motion of members in the L1641 Cloud and positioned V1787 Ori A \& B in the vector point diagram (VPD) for establishing their membership with the L1641 cloud. The stars originating from the same population such as open clusters \citep{Strai2019A&A...623A..22S}, star-forming regions \citep{Best2017ApJ...837...95B} and Nearby Young Moving Groups \citep{Ujjwal2020AJ....159..166U} are identified to be clustered in the VPD because of their similar movement through space. 
\autoref{Fig2} shows the VPD including V1787 Ori A \& B and the low and intermediate-mass members of L1641 \citep{Hsu2012ApJ...752...59H, Hsu2013ApJ...764..114H}. It is seen that V1787 Ori A \& B are positioned well inside the clustered region of L1641 members in the VPD. This suggests that both V1787 Ori A \& B are members of the L1641 cloud, which further supports our claim that they form a binary system. However, 2MASS J05380922-0649272 (which was the third star used in our binary analysis) is positioned farther away from the L1641 population. This suggests that 2MASS J05380922-0649272 is not associated with the L1641 cloud. Also, our analysis shows that the mean distance value of 2MASS J05380922-0649272 is higher than that of V1787 Ori A \& B and hence, we confirm that it is indeed a background star without any association with the V1787 Ori system.

\subsection{VLT/NACO Adaptive Optics Imaging}

We used the adaptive optics (AO) imaging data taken using Nasmyth Adaptive Optics System (NAOS) Near-Infrared Imager and Spectrograph (CONICA) (in short NAOS-CONICA or NACO) for the present study. NACO was an AO instrument at the VLT which operated in imaging, imaging polarimetry, and coronagraphy in the range of 1$-$5 micron \citep{NACO_A, NACO_B, Messenger}. It was initially deployed at the Nasmyth B focus of UT4 from 2001 to 2013 and was later moved to Nasmyth A focus of UT1 from 2013. It was decommissioned from October 2019. The NACO archival data in K\textsubscript{s} filter is available for V1787 Ori, taken on 22-03-2008 and 25-02-2008. We used the former dataset for our present study. The observation on 25-02-2008 is not of good quality compared with the data taken on 22-03-2008 and is not useful to our study. The NACO K\textsubscript{s} image of the V1787 Ori region was processed using the calibration files such as master dark and flats provided in the ESO science archive. 

\begin{figure}
   \centering
   \includegraphics[width=1\columnwidth]{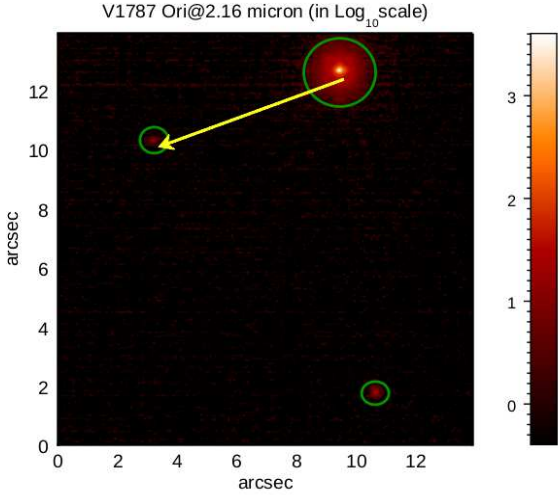}
   \caption{The processed NACO K\textsubscript{s} band image of V1787 Ori and the 2 nearby stars is shown. V1787 Ori A is shown in the larger green circle. The yellow arrow is pointing towards the wide binary companion V1787 Ori B.}
   \label{Fig3}
   \end{figure}

The AO observations of V1787 Ori A \& B were performed in K\textsubscript{s} filter in jitter imaging mode with integration time (DIT) of 1.5~s. The number of frames (NDIT) is 20, with a pixel sampling of 27 mas over a period of 3.3 minutes. The airmass corrected atmospheric seeing retrieved from the header (TEL IA FWHM) was 0.92 arcsec. The atmospheric coherence time was 5 to 7 ms. The images were dark subtracted, flat fielded, and then corrected for bad pixels using IDL software. A mean Strehl ratio of 0.2 was estimated from the AO corrected images of V1787 Ori A, which is good under the given atmospheric conditions. We estimated the instrumental magnitudes of V1787 A \& B from these pre-processed images using three different methods described below. The processed image is shown in \autoref{Fig3}. The stars encircled in green are V1787 Ori A, B, and 2MASS J05380922-0649272. 

In the first method, 128$\times$128 pixel segments of the frames centered at V1787 Ori A were obtained, and aperture photometry was done on each of them.  Although the image exhibited a diffraction-limited core, an extended halo was seen over a 40$\times$40 pixels region centered on the target. The radius of the aperture (22, 24 pixels) was selected carefully by visual inspection to include the flux in the halo as well. The estimated flux was converted to magnitudes after doing the aperture correction. Finally, an average magnitude was estimated from five object frames. A similar procedure was followed on V1787 Ori B with a segment size of 64$\times$64 pixels. The aperture radius (16 pixels) was again chosen after visual inspection of the image displayed on the logarithmic scale. The aperture correction estimated from the brighter target V1787 Ori A was used for V1787 Ori B as well. In both cases, choosing the aperture center as the center of a model Gaussian fit with sub-pixel accuracy (as against choosing the pixel with a maximum count as the center) did not make any significant difference in the magnitude estimate. In the second method, 128$\times$128 pixel segments of the images (64$\times$64 for V1787 Ori B) centered on the targets were registered with a tenth of a pixel accuracy using 2D cross-correlations and a median image was obtained. Magnitudes were estimated from the median combined images using aperture photometry described in the previous method.
In the third method, adjacent frames were subtracted from each other and a set of eight difference image frames were obtained. This procedure eliminates the sky background provided, and it does not change significantly over the 50 s interval between the frames. Magnitudes were estimated again from the expected positions of the targets using aperture photometry described earlier. Finally, an average magnitude was estimated after carefully omitting the outliers.

The mean instrument magnitude (ZP(K\textsubscript{s})) estimated from the standard star measurements is 20.1579 ± 0.0558. Also, adopting the standard value of 0.065 mag/airmass published by \cite{Nikolaev2000} for the atmospheric absorption, the magnitude correction is 0.143. Thus, we estimate K\textsubscript{s} magnitudes of V1787 Ori A \& B to be 7.75 $\pm$ 0.073 mag and 13.05 $\pm$ 0.116 mag, respectively \footnote{We noticed that there is slight difference between NACO K\textsubscript{s} and VISTA K\textsubscript{s} magnitudes from \cite{Meingast2016}. This could be due to (a) inaccurate high order correction arising from angular anisoplanatism. This implies that the error due to anisoplanatism could be as large as 0.4 magnitudes. (b) the dynamical evolution of the system. This could be proved only with multi-epoch imaging of the system.} 

The angular separation between V1787 Ori A \& B was estimated from the processed images as $6.66\pm0.02\arcsec$. This translates into a projected distance of 2577 $\pm$ 1 au between them, given that the distance of HAeBe star V1787 Ori A is 387$\pm$8 pc \citep{bailerjohns2018}. Such a large separation between the stars includes the V1787 Ori system to the class of wide binaries. 

\subsection{Spectral type estimation of V1787 Ori B}\label{sec:spty}

Having identified V1787 Ori B to be associated with V1787 Ori A, the next step is to estimate the stellar parameters. Estimation of spectral type is very important in characterizing a star, particularly if it belongs to a binary system. The best and accurate way to estimate the spectral type is from the optical/infrared spectrum of the star. A search through the archival databases such as LAMOST and APOGEE did not provide any spectral data on V1787 Ori B. The alternative method is to use the available photometric data to estimate the spectral type. In the following, we estimate the spectral type of V1787 Ori B in two different ways: (i) from an analysis of optical and infrared photometric colors and (ii) from template match to the spectral energy distribution.

\subsubsection{Spectral type from photometric colors}\label{sec:colors}

 The photometric data from SDSS, Pan-STARRS, and Vienna
survey in Orion (VISION) survey programs are used for the estimation of spectral type. We used the color--spectral type look-up tables listed in \cite{WEST2005}, \cite{Best2018} and \cite{mamajeck2013} for the conversion of SDSS (i$-$z), Pan-STARRS (i$-$z) and 2MASS (H$-$K\textsubscript{s}) colors to the corresponding spectral types. The Gaia DR2 color$-$spectral type look-up table from \cite{Kiman2019} is not used because the BP/RP excess factor, indicative of the light contamination in the photometry, is 2.6 for V1787 Ori B. For good quality Gaia photometry BP/RP excess factor should be nearly unity \citep{Riello2018}. Also, it is to be noted that for V1787 Ori B, the 2MASS magnitude photometric confusion flag (cc\_flg)\footnote{https://old.ipac.caltech.edu/2mass/releases/allsky/doc/sec4\_7.html} is shown to be "C". This means that the photometry is contaminated by a nearby star. Hence, we took VISION magnitudes, calibrated to the 2MASS photometric system, for the near-IR analysis of V1787 Ori B. 
 
 The extinction towards V1787 Ori B is not reported in any previous studies. We employed the following method to estimate the extinction (A\textsubscript{V}) value of V1787 Ori B. We used a grid of extinction values from A\textsubscript{V} = 0 to A\textsubscript{V} = 4, in steps of 0.5 mag. The maximum line of sight extinction value of A\textsubscript{V} = 4 is set due to V1787 Ori A \citep{Vioque2018}. Considering the fact that both the stars are very close to each other($\sim$ 6") and at a similar distance, the maximum extinction A\textsubscript{V} = 4 is a fairly good assumption for V1787 Ori B as well. The grid of A\textsubscript{V} values is extinction corrected using \cite{Mathis1990} extinction law, and the equivalent spectral types are listed in \autoref{tab:my-table2}.  It can be seen that as the extinction correction increases, the spectral types start to converge to a common type in all the three different colors. At A\textsubscript{V} = 4, the spectral type is converged to M3$-$M4 from the analysis of SDSS (i$-$z), Pan-STARRS (i$-$z) and 2MASS (H$-$K\textsubscript{s}) colors. This convergence of spectral type at high A\textsubscript{V} value indicates that V1787 Ori B has an extinction similar to V1787 Ori A.
 
 \begin{table*}
\centering
\caption{The table lists the spectral type estimation of V1787 Ori B using SDSS, Pan-STARRS and 2MASS colors, which are extinction corrected from 0 to 4 in steps of 0.5. The last two columns shows the $\chi^2$ values and resultant spectral type from the SpeX spectral fitting done on the SED of V1787 Ori B.}
\label{tab:my-table2}
\resizebox{\textwidth}{!}{%
\begin{tabular}{|c|c|c|c|c|c|c|c|c|}
\hline
\begin{tabular}[c]{@{}c@{}}A\textsubscript{V}\\ (mag)\end{tabular} & \begin{tabular}[c]{@{}c@{}}SDSS\\ (i$-$z)\\ (mag)\end{tabular} & \begin{tabular}[c]{@{}c@{}}Spectral Type\textsuperscript{a}\\ \end{tabular} & \begin{tabular}[c]{@{}c@{}}Pan-STARRS\\ (i$-$z)\\ (mag)\end{tabular} & \begin{tabular}[c]{@{}c@{}}Spectral Type\textsuperscript{b}\\ \end{tabular} & \begin{tabular}[c]{@{}c@{}}2MASS\textsuperscript{*}\\ (H$-$K\textsubscript{s})\\ (mag)\end{tabular} & \begin{tabular}[c]{@{}c@{}}Spectral Type\textsuperscript{c}\\ \end{tabular} & \begin{tabular}[c]{@{}c@{}}SpeX library fit\\$\chi^2$\\ \end{tabular} & \begin{tabular}[c]{@{}c@{}}Spectral Type\textsuperscript{d}\\ \end{tabular}\\ \hline
0 & 1.59 & M8 & 1.1946 & M8 & 0.521 &  L3 & 1.91 & L4 \\ 
0.5 & 1.4911 & M8 & 1.1160 & M8 & 0.49 & L2-L3 & 2.75 & L1.5\\ 
1 & 1.3922 & M7 & 1.0375 & M7 & 0.459 & L2 & 2.92 & M9\\ 
1.5 & 1.2933 & M7 & 0.9590 & M6 & 0.428 & L2 & 3.5 & M9\\ 
2 & 1.1944 & M7 & 0.8805 & M5 & 0.397 & L1 & 2.9 & M8\\ 
2.5 & 1.0955 & M6 & 0.8019 & M5 & 0.366 & M9 & 2.5 & M6\\ 
3 & 0.9966 & M5 & 0.7234 & M4 & 0.335 & M8 & 1.83 & M6\\ 
3.5 & 0.8977 & M4 & 0.6449 & M4 & 0.304 & M7 & 1.7 & M5 \\ 
4 & 0.7988 & M3$-$M4 & 0.5664 & M3 & 0.273 & M3.5 & 2 & M5\\ \hline
\end{tabular}%
}
\footnotesize{a:\cite{WEST2005} b: \cite{Best2018} c: \cite{mamajeck2013} d: SpeX library best fit}\\
\footnotesize{\textsuperscript{*} 2MASS magnitudes are H \& K\textsubscript{s} magnitudes from \citep{Meingast2016}}

\end{table*}  

\subsubsection{Spectral type from spectral template fit to the SED}

\begin{figure}
   \centering
   \includegraphics[width=1\columnwidth]{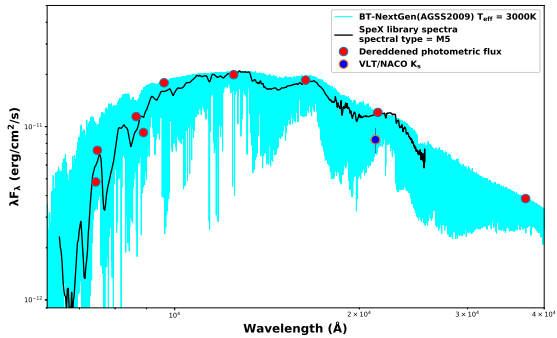}
   \caption{Figure illustrates the least $\chi^2$ SED template fit using SpeX library for the star V1787 Ori B. The fit gives an extinction value, A\textsubscript{V} = 3.5 and a spectral type as M5. The fitted SED is overplotted with BT-Next Gen(AGSS2009) synthetic spectra of T\textsubscript{eff} = 3000~K}
   \label{Fig4}
\end{figure}

 The spectral energy distribution (SED) of V1787 Ori B is constructed using magnitudes in various passbands, using Virtual Observatory SED Analyzer (VOSA) \citep{Bayo2008}. Since V1787 Ori B is a faint M-type star, we did not find good quality $B$, $V$, $G$, G\textsubscript{BP}, and G\textsubscript{RP} magnitudes in the literature. Hence, we used SDSS $i$, $z$, Pan-STARRS $i$, $z$, $y$, VISTA $J$, $H$, $K\textsubscript{s}$, VLT/NACO $K\textsubscript{s}$ and L' \citep{Connelley2008} magnitudes for generating the SED. This SED is matched with templates corresponding to various spectral types from SpeX Prism spectral library \citep{SpeX2014}. The spectral type and $\chi^2$ values obtained from the template match with the SED for A\textsubscript{V} = 0 to 4 are given in the last two columns of \autoref{tab:my-table2}. The spectral types range from L4 to M5 when A\textsubscript{V} changes from 0 to 4. The $\chi^2$ values from the template match seems to be lowest for A\textsubscript{V} = 3.5, corresponding to a spectral type of M5. Thus, the best estimates of spectral type and A\textsubscript{V} for V1787 Ori B are taken as M5 spectral type and A\textsubscript{V} = 3.5. The SED of V1787 Ori B corrected for A\textsubscript{V} = 3.5, overplotted with SpeX library spectra of spectral type M5 and BT-Next Gen (AGSS2009) synthetic spectra of T\textsubscript{eff} = 3000~K is shown in \autoref{Fig4}. 

The spectral type and extinction value of V1787 Ori B derived from this method are consistent with the color index--spectral type conversion method explained in Sect. \ref{sec:colors}. For A\textsubscript{V} values between 3.5 and 4, the spectral type of V1787 Ori B obtained from photometric colors is in the range $M3-M7$. Hence, combining the analysis from both the methods we conclude that the spectral type of V1787 Ori B is M5 $\pm$ 2. 
 
\begin{figure*}
   \centering
   \includegraphics[width=1\columnwidth]{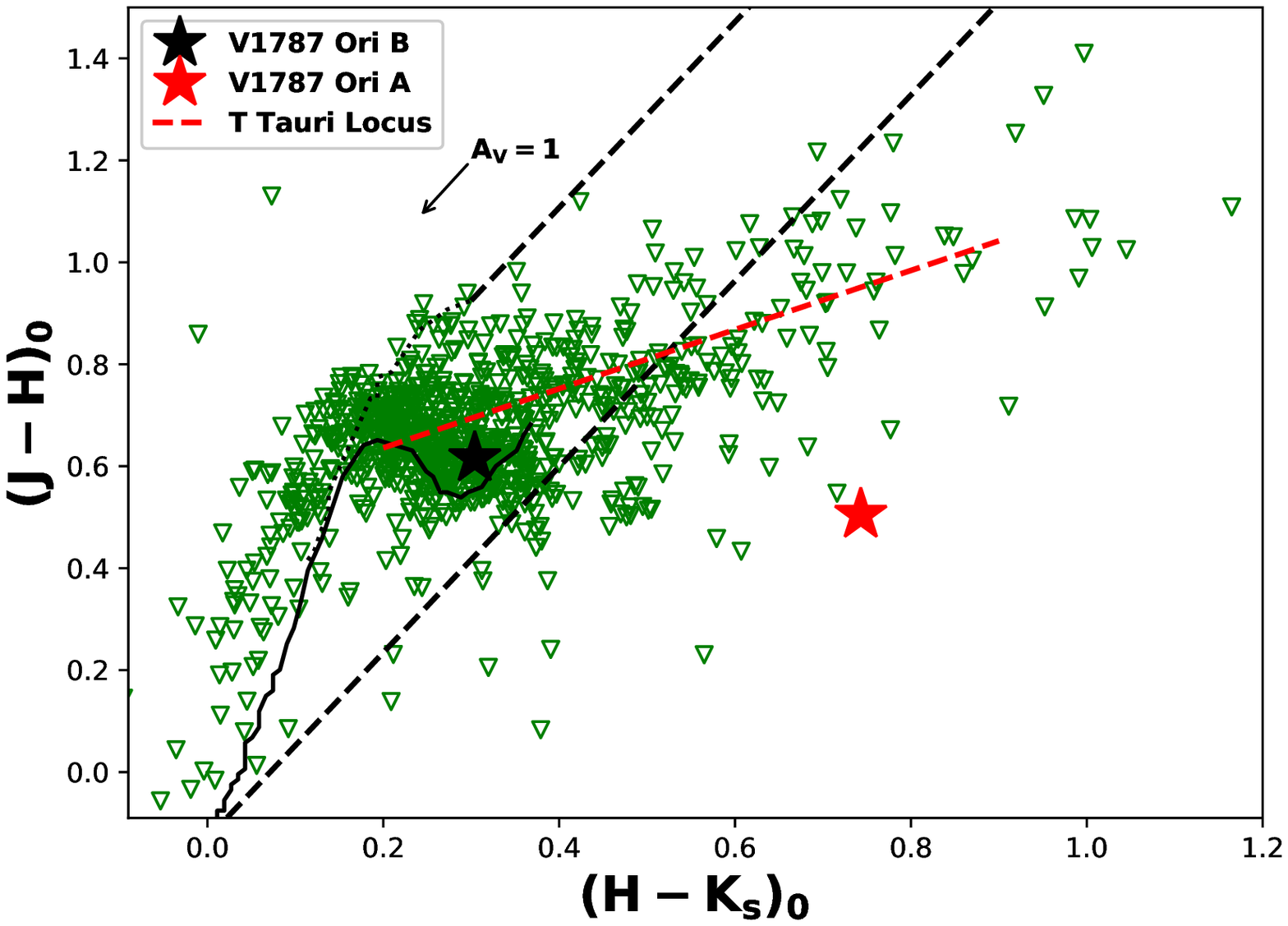}
   \includegraphics[width=1\columnwidth]{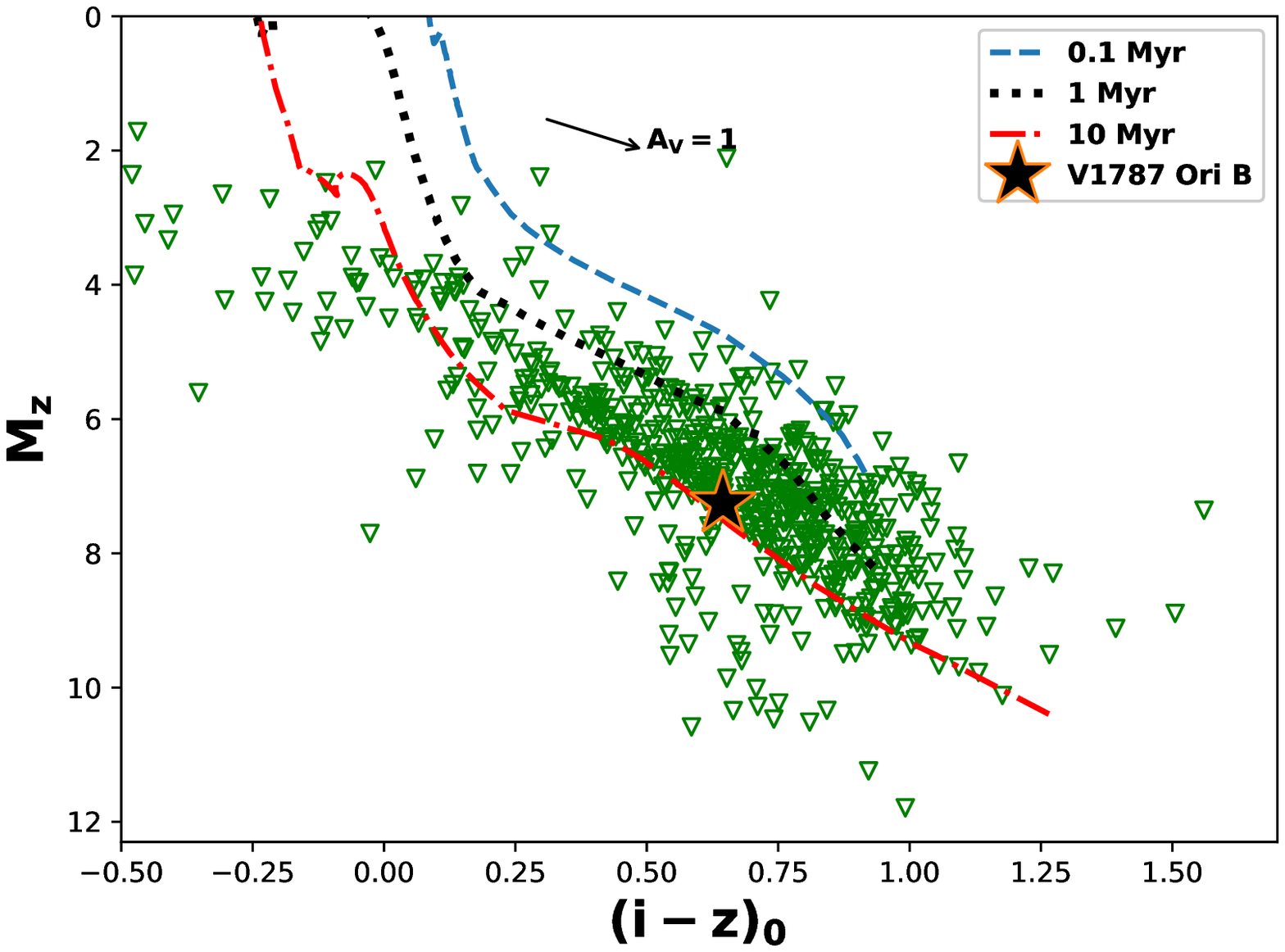}
\caption{Figure shows JHK\textsubscript{s} CCDm of V1787 Ori A \& B (left) and Pan-STARRS CMD of V1787 Ori B (right). The background stars in both diagrams are YSOs from L1641 taken from \protect\cite{Fang2013}. V1787 Ori B is denoted in both the diagrams using the black star symbol. The JHK\textsubscript{s} CCDm is included with the main sequence and giant sequence tracks adopted from \protect\cite{Koorneef1983} and converted to the 2MASS system using \protect\cite{Carpenter2001}. The V1787 Ori B is positioned inside the reddening vector and does not show near-IR excess, which indicates that the inner disc has already dissipated. The PARSEC models are overplotted on the CMD to show the youthness of V1787 Ori B. The age and the mass estimated for the star is 8.1$^{+1.7}_{-1.5}$ Myr and 0.39$^{+0.02}_{-0.05}$ M\textsubscript{\(\odot\)} respectively.} 
   \label{Fig5}
\end{figure*} 

\subsection{Infrared flux excess from near-IR color-color diagram}\label{sec:extinction}

The position of a star in the near-IR $(J-H)$ vs $(H-K\textsubscript{s})$ color-color diagram (CCDm) is used to investigate whether protoplanetary disc in present in PMS stars. The dust in the disc emits in near-IR and hence shows flux excess in the near-IR colors $(J-H)$ and $(H-K\textsubscript{s})$. This shifts the location of PMS stars such as T Tauri and Herbig Ae/Be stars to redder colors in $(J-H)$ and $(H-K\textsubscript{s})$. The extinction towards a star is the prime factor affecting its position in the CCDm. The near-IR CCDm representing V1787 Ori A \& B is shown in \autoref{Fig5}. The CCDm also includes Young Stellar Objects (YSOs) from L1641, reported in \cite{Fang2013}, as a reference sample. The YSOs are corrected for reddening using the A\textsubscript{V} reported by \cite{Fang2013}. The near-IR colors of V1787 Ori A are corrected using A\textsubscript{V} = 4 while that of V1787 Ori B is corrected using A\textsubscript{V} = 3.5, as estimated from Sect. \ref{sec:spty}. The main sequence and giant sequence are adopted from \cite{Koorneef1983}. The T Tauri locus from \cite{Meyer1997} is also indicated in the CCDm. While representing in \autoref{Fig5}, the intrinsic colors of stars in these sequences are converted to the 2MASS system using the transformation relations in \cite{Carpenter2001}. As evident from the classification, the Herbig Ae star V1787 Ori A show near-IR excess, indicative of the presence of hot dust in the circumstellar disc. As seen from \autoref{Fig5}, V1787 Ori B is located inside the reddening vector and does not show near-IR excess, which indicates that the inner disc may have dissipated.

\subsection{Age and mass of V1787 Ori system from optical color-magnitude diagram}
\label{sec:age}

The age and mass of V1787 Ori B is estimated from their location in the Pan-STARRS $M\textsubscript{z}$ vs $(i-z)$ color-magnitude diagram (CMD). The CMD is extinction corrected using A\textsubscript{V} = 3.5 for V1787 Ori B and is shown in \autoref{Fig5}. The reference sample of YSOs from \cite{Fang2013} which are illustrated in near-IR CCDm is also represented in the CMD. The L1461 YSO catalog is cross-matched with Pan-STARRS and the stars with i and z magnitude uncertainity less than 0.02 mag are plotted in the CMD. The CMD is overlayed with isochrones of ages 0.1, 1 and 10 Myr from PARSEC\footnote{http://stev.oapd.inaf.it/cgi-bin/cmd} \citep{Bressan2012MNRAS.427..127B,Marigo2017ApJ...835...77M}. From the location in the CMD (see \autoref{Fig5}), the age of V1787 Ori B is estimated to be 8.1$^{+1.7}_{-1.5}$ Myr and the mass is estimated as 0.39$^{+0.02}_{-0.05}$ M\textsubscript{\(\odot\)}. 

As mentioned in the previous section, V1787 Ori A \& B is identified to be part of the L1641 molecular cloud. The YSOs in L1641 has a median age of $\sim$ 1.5 Myr \citep{Fang2013}. But there are stars with ages more than 10 Myr. The age estimated for V1787 Ori B from this study is consistent with the older generation of stars from the L1641 cloud. The binary system that formed together should be coeval \citep{Kraus2009}. Hence, the age of V1787 Ori A should be matching with the age of V1787 Ori B. The quality of Pan-STARRS i \& z magnitudes of V1787 Ori A is poor, which restricts us from consistently estimating the age of both the stars from $M\textsubscript{z}$ vs $(i-z)$ CMD. The ages reported for V1787 Ori A by \cite{Arun2019} and \cite{Vioque2018} are 0.50$\pm$ 0.01 Myr and 7.4$^{+0.6}_{-1}$ Myr respectively. The difference in the age values of V1787 Ori A in these studies is majorly due to two parameters. First one is the different total-to-selective extinction (R\textsubscript{V}) values employed while calculating the extinction and the second is the models used for estimating the age. The age estimated for HAeBe stars by \cite{Vioque2018} used R\textsubscript{V} = 3.1, which is considered as the average interstellar R\textsubscript{V} value. Also, the study used the PARSEC models. However, \citet{Gorti1993}, \citet{Hernand2004} and \cite{manoj2006} have mentioned that, for HAeBe stars, due to the presence of larger grains in the circumstellar environment, can have a higher R\textsubscript{V}. And the generally adopted value in studies of HAeBe stars is R\textsubscript{V} = 5.  Hence, the study by \cite{Arun2019} dealing with a population of HAeBe stars, adopted R\textsubscript{V} = 5 in the work. Also, \cite{Arun2019} used the Modules for Experiments in Stellar Astrophysics (MESA) isochrones and evolutionary tracks (MIST) \citep{choi2016, Dotter2016}. Since, we are considering the V1787 Ori system individually, we estimated  the R\textsubscript{V} value independently for V1787 Ori A. We used the equation given by \cite{Cardelli1989ApJ...345..245C} and \citep{manoj2006} for the estimation and the value is found to be R\textsubscript{V} $\sim$ 3.1. This makes the age estimates of \cite{Vioque2018} more suitable for V1787 Ori A. The age of V1787 Ori B from Pan-STARRS CMD is consistent with the age of V1787 Ori A from \cite{Vioque2018}. Considering the models used for age estimation and calculation of R\textsubscript{V} value of V1787 Ori A, age value of V1787 Ori A from \cite{Vioque2018} is given preference in this study, making the V1787 Ori system coeval. 

To summarize, from the analysis of optical CMD and near-IR CCDm, we found that V1787 Ori B lacks IR excess and is in the PMS phase with a mass of 0.39$^{+0.02}_{-0.05}$ M\textsubscript{\(\odot\)} and age of 8.1$^{+1.7}_{-1.5}$ Myr. The age estimate of V1787 Ori A from \cite{Vioque2018} is 7.4$^{+0.6}_{-1}$ Myr making the binary system coeval. 

\section{Discussion}\label{sec:disc}

\subsection{Probability analysis of the chance alignment of components}

The binarity of V1787 Ori A \& B is confirmed by the Gaia DR2 astrometric analysis. The analysis showed that the distance and proper motion values of both stars match within uncertainties and they kinematically belong to L1641 cloud. However, as mentioned earlier, proper motion uncertainties are very high. Thus, we have to check the probability of chance alignment of the two stars in the V1787 system to support our claim of the association of V1787 Ori A \& B.

To investigate whether the two stars are associated with each other, we calculated the probability of such a system arising naturally due to the two random stars being close to each other. To do so, we used the Cross\_prob code \citep{cross_prob}, the steps of which are explained further. First, we start with the stellar number densities for M- and A-type stars per pc\textsuperscript{3}. We used the number density of M dwarfs as 0.09 pc\textsuperscript{-3}, taken from \cite{Kirkpatrick2012}, and for A-type stars the number density\footnote{http://www.pas.rochester.edu/~emamajek/memo\_star\_dens.html} is 5$\times$10$^{-4}$ pc\textsuperscript{-3}.  We then calculated the number density for M- and A-type stars at a distance of 390 pc in a shell of 100 pc ($\sim$ twice the error in the uncertainty of V1787 Ori B). This resulted in 426 M-type stars and 2 A-type stars in the shell per degree\textsuperscript{2}. Since V1787 Ori A \& B are separated by 6.7$\arcsec$, we simulated the occurrence of A-type stars within  6.7$\arcsec$ of an M star by repeatedly creating a fiducial distribution of M- and A-type stars and counting the instances when an A-type star was within 6.6$\arcsec$ of M-type stars. Based on 100,000 such independent iterations, the probability of an A-type with 6.6$\arcsec$ of M-type stars is 1.0 $\pm$ 0.1 \%. This means that the probability of a chance alignment of an A-type star and M-type star is $\sim$1\%, which is very low. Thus, V1787 Ori A \& B is much more likely to be a bound pair. 

\begin{table*}
\caption{Table lists the magnitudes of V1787 Ori A \& B in various passbands used for this study. The NACO K\textsubscript{s} magnitude of the stars are estimated from this work.}
\label{tab:my-table3}
\resizebox{\textwidth}{!}{%
\begin{tabular}{cccccccccc}
\hline
Star & J$^*$ & H$^*$ & K\textsubscript{s}$^*$ & L'$^1$ & i$^2$ & z$^2$ & i$^3$ & z$^3$ & NACO K\textsubscript{s} \\ \hline
V1787 Ori A & 9.938$\pm$0.021 & 8.969$\pm$0.027 & 7.978$\pm$0.027 & 6.15$\pm$0.05 & 12.593$\pm$0.003 & 11.980$\pm$0.005 & 12.7700 & 11.9610 & 7.75$\pm$0.073 \\
V1787 Ori B & 14.168$\pm$0.006 & 13.146$\pm$0.004 & 12.625$\pm$0.005 & 12.25$\pm$0.08 &  17.864$\pm$0.010 & 16.805$\pm$0.013 & 18.1183$\pm$0.0143 & 16.9237$\pm$0.0166 & 13.050$\pm$0.116 \\ \hline
\end{tabular}%
}
\footnotesize{1: \cite{Connelley2008}, 2: SDSS, 3: Pan-STARRS}\\
\footnotesize{\textsuperscript{*} JHK\textsubscript{s} magnitudes are VISION VISTA magnitudes from \cite{Meingast2016}}
\end{table*}

\subsection{On the binarity of V1787 Ori system}

The HAeBe star V1787 Ori A is not reported as a binary system in previous studies \citep{Connelley2008, Vioque2018}. The wide binary companion of V1787 Ori A is identified as an M5$\pm$2 type star at a projected separation of 2577 au. The age and mass of V1787 Ori B are found to be  8.1$^{+1.7}_{-1.5}$ Myr and 0.39$^{+0.02}_{-0.05}$ M\textsubscript{\(\odot\)} using PARSEC stellar model. We adopt the age of V1787 Ori A to be 7.4$^{+0.6}_{-1}$ Myr \citep{Vioque2018} in this study (as seen from Sect. \ref{sec:age}).  Thus, it is safe to assume that V1787 Ori A \& B are coeval. However, to constrain the uncertainty in the stellar parameters, consistently measured good quality photometric and spectroscopic data for both the stars need to be obtained, which is beyond the scope of this work.

V1787 Ori is a PMS wide binary system with component A being a HAeBe star and component B a low mass (M-type) PMS star. Since we do not have spectroscopic information for V1787 Ori B, we are unable to confirm whether H$\alpha$ emission is present and hence not sure whether the star is accreting. However, since the star is young and in the PMS phase (as deduced from CMD), and with similar $(J-H)$, $(H-K)$ colors as that of low mass YSOs, we suggest V1787 Ori B to be a low mass PMS star. The study by \cite{Bouvier2001} on Herbig visual binary systems suggested that the disc lifetime of the low mass companion can be shortened by the influence of the high mass primary. The lack of infrared excess in V1787 Ori B, as seen from the SED and the near-IR CCDm, can be due to the effect of V1787 Ori A..

\cite{Duch2015} discussed the multiplicity in HAeBe stars and mentioned that about 90\% of the HAeBe stars could have a binary companion. The binarity and mass ratios of HAeBe stars are studied by \cite{Baines2006} and \cite{Wheel2010} using spectroastrometry. Even though both the studies have completeness issues, the binarity found by both the works is similar (68$\pm$11\% \& 74$\pm$6\%). The findings by \cite{Wheel2010} suggest that most of the identified HAeBe star binaries should have a high mass ratio ({\it q} = 0.7). The mass of V1787 Ori A is reported to be 1.66  M\textsubscript{\(\odot\)} \citep{Vioque2018}. Using the mass of V1787 Ori B estimated from this study, the mass ratio of the V1787 Ori system is estimated as 0.23. This makes the V1787 Ori binary system rarely identified candidate among the sample of HAeBe binaries. Interestingly, Volume-limited A-Star (VAST) survey by \cite{VAST1} studied the multiplicity of A-type stars including main sequence and PMS stars in the near solar neighborhood. The survey identified 137 A-type stars having companions \citep{VAST32014}. The study shows that even though the {\it q} values suggested to be preferred by HAeBe stars are slightly at a higher-end \citep{Wheel2010}, the {\it q} $\sim$ 0.1 are preferred in the case of A-type stars in the solar vicinity. Also, a lower distribution of {\it q} values are observed in the binaries reported among intermediate mass stars in the Scorpius OB2 association \citep{Kouwenhoven2007A&A...474...77K}.

The observed higher $q$ values in HAeBe stars maybe due to the observational bias created by the sensitivity of the instruments used till now. The new and advanced instruments such as VLT/SPHERE can mitigate this issue and can detect fainter companions. The detection rates of low mass companions to YSOs are increasing in recent times. The recent detection of a low mass companion to MWC 297 using VLT/SPHERE \citep{Giulia2020} and a second companion to the GQ Lup system using Gaia DR2 \citep{GQLup2020} are examples worth mentioning. More concentrated studies on PMS stars with the advanced instruments can solve the completeness issues in the multiplicity studies of these stars. Other reported cases of M-type companion to HAeBe stars are that of HD 142527 \citep{Lacour2016} and HD 100453 \citep{Dong2016}. In both these cases, however, the companions are nearby companions, unlike in the case of V1787 Ori A. A few noteworthy studies on wide binary companions of the PMS stars are Herbig Ae-Ae wide binary system PDS 144 \citep{Hornbeck2012}, T-Tauri wide binary system PDS 11 \citep{Mathew2017} and TW Hya--brown dwarf binary system \citep{Teixeira2008}. 

\subsection{On the possible binary system formation scenarios}

The available data is not enough to identify the formation scenario of the V1787 Ori binary system. However, in this section, we discuss all the possible binary formation scenarios and assess which one suits for V1787 Ori. The monolithic collapse of a core and subsequent disc fragmentation was proposed by \cite{Wheelwright2011} for the formation of HAeBe binaries. This mechanism is preferred for near-equal mass binaries \citep{Krumholz2009} and this is true for the case of HAeBe binaries as they have high {\it q} value of $\sim$ 0.7. Another proposed scenario is disc fragmentation scenario that can form extreme mass ratio binary with a low mass or brown dwarf secondary star for solar mass stars are proposed by \cite{Stamatellos2009}. But the separation of V1787 Ori system is very high for this to be possible. Also, there is reported case of disk fragmentation at a distance of $\sim$ 2000 au for a massive O type protostar with extreme mass ratio \citep{Ilee2018}. An A type star having such an extended disc is less likely. V1787 Ori system has a $q$ value of $\sim$ 0.23 with a separation of 2500 au. Such a system may not prefer monolithic collapse and a disk fragmentation. The massive star formation can also take the course of competitive accretion scenario \citep{Bonnell2005}. In competitive accretion, the stars form and accrete from the same parent cloud and due to the stellar capture, they come close to each other \citep[e.g. M17-][]{Sana2017}. In the competitive accretion scenario, both the stars will be close to each other and the system has a high $q$ ratio, which makes it not a viable formation scenario for the V1787 Ori system.  

A distant, but possible scenario, is the formation of binary by capture \citep{Bodenheimer2011}. A three-body capture can produce a very wide binary companion, but it can happen in very dense young star-forming environments or happens only in rare instances in our Galaxy \citep{Mansbach1970, Boss1988}. Also, a two-body capture could only produce a very close binary system \citep{Bodenheimer2011}. The rarest but possible capture scenario is the dissipative capture, where the protostellar envelope or the circumstellar medium could act as a dissipative medium in capturing the possible companion star \citep{Boss1988}. This scenario also cannot explain the formation of the V1787 Ori binary system since the companion is in a wide orbit at a separation of 2577 au. V1787 Ori binary system is found to belong to the L1641 cloud whose stellar surface density is $\sim$10 stars pc\textsuperscript{-2}. Dissipative capture may not work in such a low stellar density environment. Another process that is proposed to form young multiple systems is prestellar core collapse \citep{Goodwin2005}. In this process the prestellar core form multiple stars that accrete from the same parent cloud. \cite{Goodwin2005} suggests that 2 to 3 stars could form from the same cloud through this process. The prestellar core collapse is possible for the V1787 Ori system as it can account for the low mass ratio and wide separation between the components.

A secular process that is reported to form wide companions is ejection/unfolding from a triplet system. The triplet system can unfold in time scales of millions of years and a companion can be ejected out to a wider orbit \citep{Reipurth2012Natur.492..221R}. There is no concrete evidence to suggest an inner companion with the available data. With NACO/Ks AO imaging, we do have two pixels diffraction limit of 55 mas. However, it should be noted that AO corrected PSFs (FWHM) are not always diffraction-limited owing to the residual atmospheric turbulence effects (i.e. the correction is close to perfect but not perfect). In our case, V1787 Ori A has an FWHM of 100-110 mas. Thus, we can rule out an inner companion with a separation of 80 au. Also, the Gaia DR2 astrometric parameter RUWE is 1.33 for V1787 Ori A, which suggests that the presence of an inner companion is less likely. A probable process that can form a wider companion is filament-like fragmentation. The stars are formed actively in the filament structure in Giant molecular clouds. This transient formation scenario \citep{Kuffmeier2019A&A...628A.112K} can form wide binaries of separation $\sim$ 1000 au (e.g. IRAS 16293-2422, \citealp{van_der_Wiel2019A&A...626A..93V}). L1641 cloud is composed of filament structures and is reported to have $\sim$ 70\% star-forming cores located on the filament structures \citep{Polychroni2013ApJ...777L..33P}. The V1787 Ori system is located on such a filament structure. The filament can act as an active star formation site and could have helped V1787 Ori system to form with such a wide separation. However, more observations, preferably optical/infrared spectroscopy and ALMA imaging of the components can provide new insights about the formation scenario by constraining the inclination, geometry, and morphology of the circumstellar discs of V1787 Ori A \& B \citep{Manara2019A&A...628A..95M}.

\section{Conclusion}
\label{sect:conclusion}
In this study, we report the detection of a wide binary companion, V1787 Ori B, to the HAeBe star V1787 Ori A. For the estimation of stellar parameters of V1787 Ori B, we used all possible data from the archive, particularly VLT/NACO images. The major conclusions from this study are given below.

\begin{itemize}
    \item We used the Gaia DR2 astrometric data for establishing the binarity of V1787 Ori A \& B. Both the stars have similar proper motions and distances within the uncertainties. The VPD analysis confirms that V1787 Ori A \& B are confirmed members of the L1641 cloud. 
    \item We found the separation between the components of the V1787 Ori binary system to be 2577 au from the analysis of VLT/NACO  K\textsubscript{s}-band images.
    \item The magnitude of V1787 Ori A \& V1787 Ori B estimated using VLT/NACO K\textsubscript{s}-band image is 7.75$\pm$0.073 mag and 13.05$\pm$0.116 mag, respectively.
    \item The spectral type of V1787 Ori B is estimated to be M5 $\pm$ 2  using color - spectral type relations and SpeX Prism spectral library fitting on the SED. The extinction value A\textsubscript{V} is calculated to be 3.5 mag.
    \item The PARSEC models are overlaid on the optical CMD to estimate the age and mass of V1787 Ori B, which are found to be 8.1$^{+1.7}_{-1.5}$ Myr and 0.39$^{+0.02}_{-0.05}$ M\textsubscript{\(\odot\)}, respectively. This is consistent with the age of V1787 Ori A ($\sim$ 7.5 Myr). The similarity between the ages of V1787 Ori A \& B confirms that they are coeval and part of a PMS binary system.
    \item The mass ratio is found to be 0.23, identifying this system as a rare one among HAeBe binaries. However, since this kind of mass ratio is seen among A-type binaries in the field and star forming regions, it is quite possible that there may be more low {\it q} binaries among PMS binaries such as V1787 Ori system. 
    \item The possible scenarios for the formation of V1787 Ori wide binary system are discussed. The most probable scenarios are prestellar core collapse and filament fragmentation. However, any one of the scenario can only be confirmed from further multi wavelength observations.  
\end{itemize}{}
\section*{DATA AVAILABILITY}
The VLT/NACO data underlying this article are available from the public ESO archive\footnote{http://www.archive.eso.org/}. The photometric and astrometric data are publicly available from the VizieR catalog\footnote{https://vizier.u-strasbg.fr/}.

\section*{Acknowledgements}

We would like to thank the anonymous referee for providing helpful comments and suggestions that improved the paper.  R.A. thanks Ujjwal and Sudheesh for their valuable suggestions throughout the course of the work. This work has made use of data from the European Space Agency (ESA) mission Gaia (https://www.cosmos.esa.int/gaia), processed by the Gaia Data Processing and Analysis Consortium (DPAC; https://www.cosmos.esa.int/web/gaia/dpac/consortium). Funding for the DPAC has been provided by national institutions, in particular, the institutions participating in the Gaia Multilateral Agreement. Also, we made use of the VizieR catalog
access tool, Simbad and Aladdin, CDS, Strasbourg, France. This publication makes use of VOSA, developed under the Spanish Virtual Observatory project supported by the Spanish MINECO through grant AyA2017-84089. VOSA has been partially updated by using funding from the European Union's Horizon 2020 Research and Innovation Programme, under Grant Agreement nº 776403 (EXOPLANETS-A) 




\bibliographystyle{mnras}
\bibliography{mnras} 





\bsp	
\label{lastpage}
\end{document}